\documentclass[%
reprint,
showpacs,
bibnotes,
amsmath,amssymb,
aps,
prl,
]{revtex4-1}

\newlength{\myfigwidth}
\usepackage{epsfig}
\usepackage{mathrsfs}
\usepackage{amssymb}
\usepackage{color}
\usepackage{latexsym}
\usepackage{subfigure}
\usepackage{marginnote}
\usepackage{dsfont}

\newcommand{\bcen}{\begin{center}}
\newcommand{\ecen}{\end{center}}
\newcommand{\btab}{\begin{tabular}}
\newcommand{\etab}{\end{tabular}}
\newcommand{\bdes}{\begin{description}}
\newcommand{\edes}{\end{description}}

\newcommand{\beq}{\begin{equation}}
\newcommand{\eeq}{\end{equation}}
\newcommand{\bea}{\begin{eqnarray}}
\newcommand{\eea}{\end{eqnarray}}

\newcommand{\half}{\frac{1}{2}}
\newcommand{\bary}{\begin{array}}
\newcommand{\eary}{\end{array}}
\newcommand{\benum}{\begin{enumerate}}
\newcommand{\eenum}{\end{enumerate}}
\newcommand{\bitem}{\begin{itemize}}
\newcommand{\eitem}{\end{itemize}}

\newcommand{\btau}{\mbox{\boldmath $ \tau $}}

\newcommand{\be} { \mbox{\boldmath $e$}}

\newcommand{\br} { \boldsymbol{r}}

\newcommand{\bM} { \mbox{\boldmath $M$}}

\newcommand{\bP} { \mbox{\boldmath $P$}}

\newcommand{\bzero} { \mbox{\boldmath $0$}}

\newcommand{\OneFour}{\mathds{1}}
\newcommand{\dou}{\partial}

\newcommand{\D}[1]{\mbox{d}{#1}} 
\newcommand{\grad}{\mbox{\boldmath $\nabla$}}

\newcommand{\mean}[1]{\langle #1 \rangle}

\newcommand{\ket}[1]{| #1 \rangle}

\newcommand{\eqn}[1] {eqn.~(\ref{#1})}

\newcommand{\Fig}[1]{Fig.~\ref{#1}}

\newcommand{\mylabel}[1]{\label{#1}}

\begin{document}
\title{Sensory organ like response determines the magnetism of zigzag-edged honeycomb nanoribbons}
\author{Somnath~Bhowmick$^1$}\email[]{bsomnath@mrc.iisc.ernet.in}
\author{Amal Medhi$^{2}$}\email[]{amedhi@physics.iisc.ernet.in}
\author{Vijay B Shenoy$^{2}$}\email[]{shenoy@physics.iisc.ernet.in}
\affiliation{$^1$Materials Research Center, Indian Institute of Science, Bangalore 560 012, India\\
$^2$Centre for Condensed Matter Theory, Department of Physics, Indian Institute of Science, Bangalore 560 012, India}
\date{\today}
\begin{abstract}
We present an {\em analytical} theory for the magnetic phase diagram
for zigzag edge terminated honeycomb nanoribbons described by a
Hubbard model with an interaction parameter $U$. We show that the edge
magnetic moment varies as $\ln{U}$ and uncover its dependence on the width $W$ of the ribbon. The physics of this owes its origin to the sensory
organ like response of the nanoribbons, demonstrating that  considerations beyond the usual Stoner-Landau theory are necessary to understand the magnetism of these systems. A first order magnetic
transition from an anti-parallel orientation of the moments on
opposite edges to a parallel orientation occurs upon doping with holes or electrons. 
The critical doping for this transition is shown to depend
inversely on the width of the ribbon.  Using variational
Monte-Carlo calculations, we show that magnetism is robust to
fluctuations. Additionally, we show that the magnetic phase diagram is
generic to zigzag edge terminated nanostructures such as
nanodots. Furthermore, we perform first principles modeling to show how
such magnetic transitions can be realized in substituted graphene
nanoribbons.
\end{abstract}
\maketitle

Interest and activity in magnetic nanostructures has been driven by their possible application in nanoelectronic/spintronic devices, with graphene based systems grabbing a significant fraction of the attention.\cite{westervelt2008} The remarkable electronic properties of graphene\cite{neto2009,Pati2011} with a Dirac like spectrum have made it a suitable candidate for many applications.\cite{bunch2005,miao2007,ponomarenko2008,Wurm2009} From a theoretical perspective, electron interactions and correlations  effects on the honeycomb lattice contain many interesting phenomena\cite{Kotov2010} including magnetism\cite{MengNat10} and superconductivity.\cite{Pathak2010}

Magnetism at the zigzag terminated edges of graphene has been studied
by first principles calculations\cite{son2006,young2006}, and by a simplified effective Hubbard model
\cite{PhysRevB.70.195122,rossier2007,bhowmick2008,PhysRevLett.101.037203,macdonald2009,ma:112504,ovy2010}
described by a hopping parameter $t$ and a site-local repulsion
$U$. Ref.~\onlinecite{bhowmick2008} showed that the magnetism in
graphene is robust to ``shape disorder'' of the 
nanostructure, while ref.~\onlinecite{macdonald2009} studied finite width graphene nanoribbons including the effects of doping. There are also studies of defect induced magnetism\cite{Zhang2007,oleg2007} and of magnetism of other nanostructures\cite{ezawa2007,rossier2007}. There are encouraging recent experimental signatures of magnetism\cite{Matte2009,Rao2011,Joly2010}, along with suggestions\cite{Kuntsmann2011} that extraneous effects such as reconstruction would render the magnetism fragile.

The origin of magnetic moment in zigzag edge terminated honeycomb
nanostructures has been attributed to  the edge
states\cite{nakada1996,wakabayashi1999,neto2009} - localized
electronic states which have  most weight at the edges and die
exponentially in the bulk.\cite{brey2006a,brey2006b,sandler2009}  These states are of topological
origin\cite{Ryu2002} and have been experimentally observed using
scanning tunneling microscopy.\cite{kobayashi2005,niimi2006} Magnetism at the edges is attributed to the Stoner
mechanism(see, e.g., \cite{Altland2006}) and is best discussed in terms of a
Landau theory.\cite{Altland2006} The ground state energy of
the system is expressed as
\beq\mylabel{eqn:LandauTheory}
 E(\bM) = a(U_c- U) |\bM|^2 + b |\bM|^4 + \ldots 
\eeq 
where $\bM$ is the magnetic order parameter, $a, b>0$ are positive
constants that depend on the microscopics, and $U_c$ is a critical
value of the on-site repulsion. For $U < U_c$, the energy is minimum
when $\bM = \bzero$, i.~e., system is non-magnetic. At $U = U_c$,
there is a quantum phase transition to the magnetic state, and for $U
> U_c$ one finds $|\bM| \sim \sqrt{U - U_c}$. Stoner theory\cite{Yosida1996}, based on linear response formulation, provides
an expression for $U_c \sim \frac{1}{g(\varepsilon_F)}$, where
$g(\varepsilon_F)$ is density of states of the bare system ($U=0$) at
the zero temperature chemical potential $\varepsilon_F$. Thus for a zigzag
edge terminated system, $U_c$ would vanish, since the density of states
$g(\varepsilon_F)$ diverges owing to the non dispersive nature of the
edge states. The Stoner-Landau theory would therefore suggest that a
zigzag edge will have spontaneous magnetization $\bM$ for {\em any} $U
> 0$, and furthermore that $|\bM| \sim \sqrt{U}$ for $U \ll t$.  

It is known\cite{bhowmick2010} that a zigzag edge
terminated nanoribbon has a highly nonlinear response akin to that of
sensory organs like eyes and ears. Their density response depends
{\em logarithmically} on the magnitude of an edge potential applied at 
the zigzag edges. In this paper we show that such a Weber-Fechner response\footnote{See http://en.wikipedia.org/wiki/Weber-Fechner\_law} of these nanoribbons plays a central role in determining their magnetism. Our work leads to a ``magnetic phase diagram'' of lightly doped nanoribbons, including analytical expressions for the width $W$ and $U$ dependence of the  magnetization, excitation gap, and the critical doping required to engender magnetic transitions. We also corroborate these results with variational quantum Monte Carlo calculations and show that the magnetism is robust to fluctuations. To the best of our knowledge, this is the first report that clearly points out that the usual Stoner-Landau theory is inadequate to understand magnetism of hexagonal lattice nanoribbons. This gains added significance in view of recent
  developments of cold atom optical
  lattices,\cite{Bloch2012} where honeycomb lattices have been realized and studied.\cite{Tarruell2012}

The simplest Hamiltonian that describes the energetics of interacting electrons in the honeycomb lattice
is the Hubbard model
\begin{equation}\mylabel{eqn:Hubbard}
H = -t\sum_{\langle ij\rangle,\sigma}c^{\dagger}_{i\sigma}c_{j\sigma}
+ Ue_0 \sum_{i} n_{i\uparrow} n_{i\downarrow} - \mu \sum_i n_{i \sigma}
\end{equation}
where $c^\dagger_{i \sigma}$ is the operator that creates an electron of spin $\sigma$ at site $i$, $n_{i \sigma} = c^\dagger_{i \sigma} c_{i \sigma}$ is the number operator, $t$ is the nearest neighbour hopping amplitude, $U e_0$ is the on-site Hubbard repulsion ($e_0 = \frac{\sqrt{3}}{2}t$ is a characteristic energy scale, $U$ is dimensionless), and $\mu$ is the chemical potential. The underlying triangular Bravais lattice  has a lattice parameter $a$, and the width of the the zigzag edge terminated nanoribbon is denoted by $W$. 

\begin{figure}
\centerline{\includegraphics[width=0.235\myfigwidth]{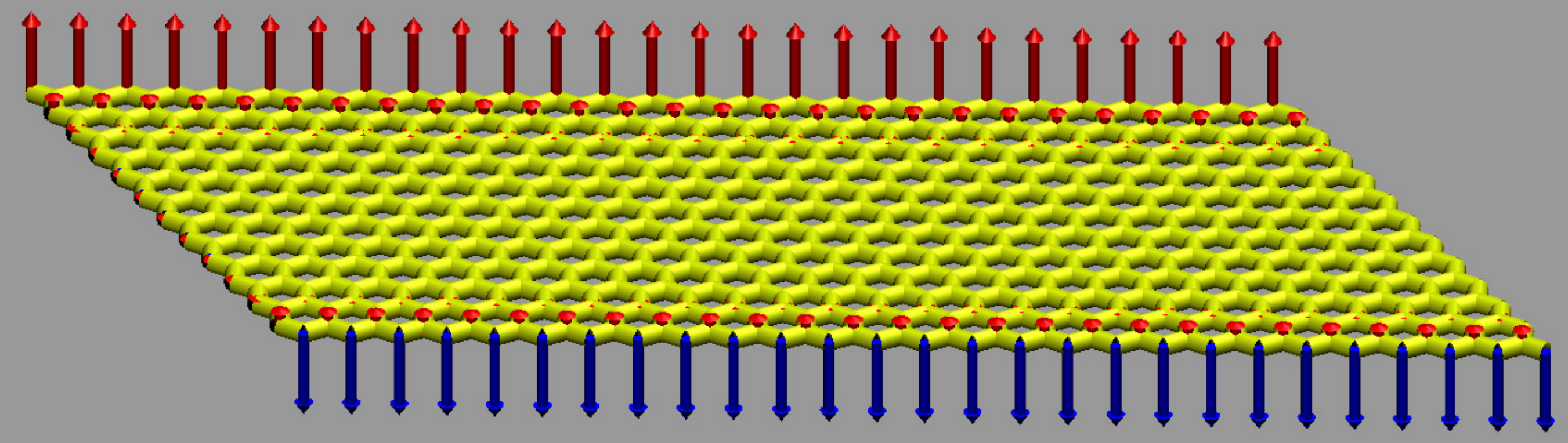}~~\includegraphics[width=0.225\myfigwidth]{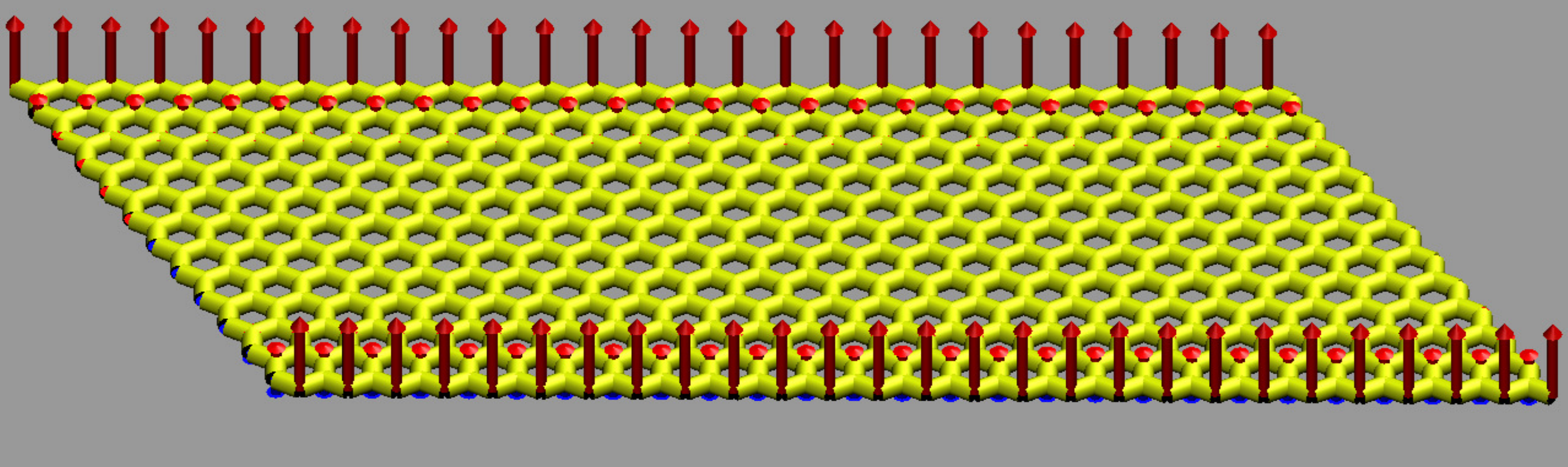}}
\centerline{(a)~~~~~~~~~~~~~~~~~~~~~~~~~~~~~~~~~~~~~~~~(b)}
\medskip
\centerline{\includegraphics[width=0.34\myfigwidth]{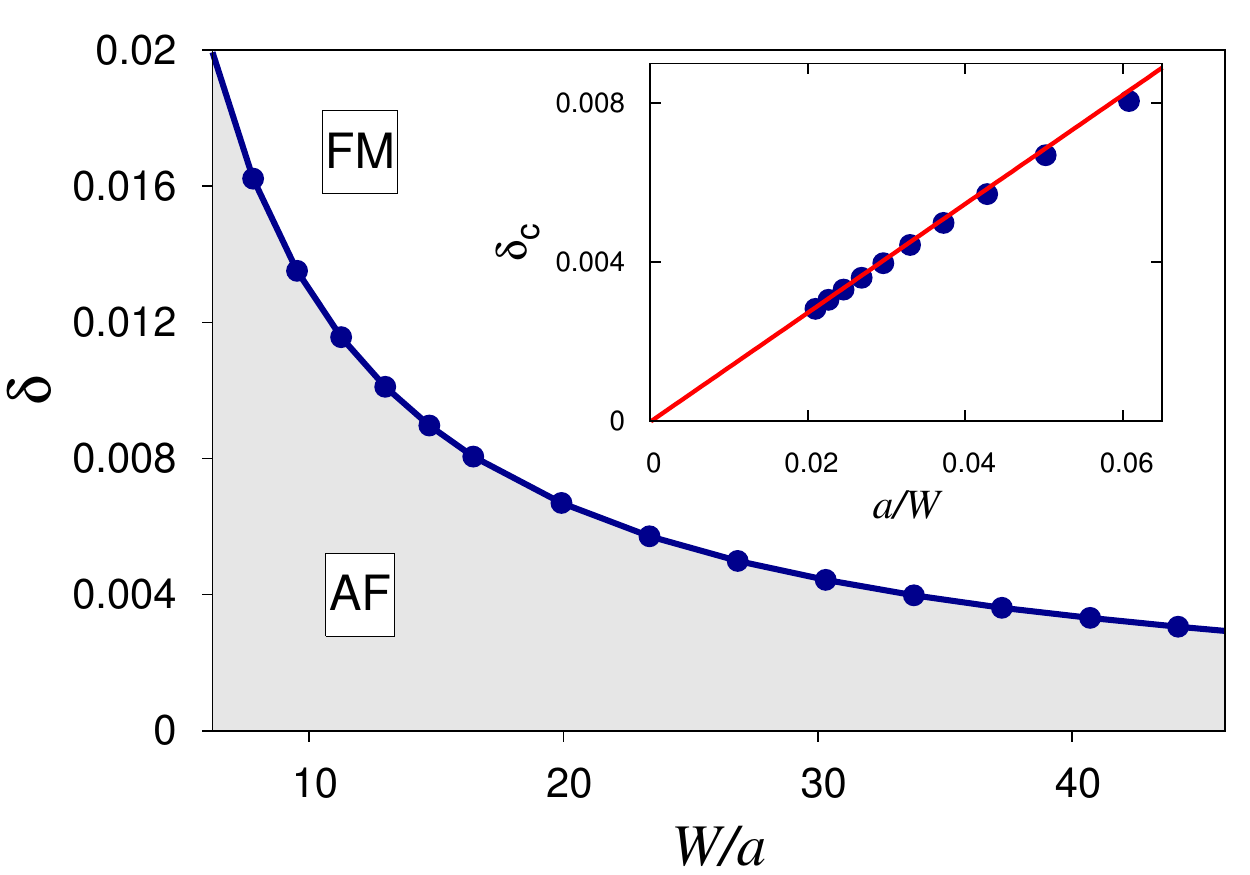}}
\centerline{(c)}
\caption{(color online) Two possible magnetic states of zigzag-edged nanoribbon,
(a) antiferromagnetic (AF) and (b) ferromagnetic (FM).
(c) Magnetic phase diagram of zigzag nanoribbons for $U=1.5$ obtained from numerical mean field calculations.
Inset: the critical doping required for AF to FM transition is inversely proportional 
to the ribbon width.}
\label{fig:maggs}
\end{figure}

Ground state of the zigzag edge terminated ribbon is obtained by a mean field analysis, where the four fermion interaction term is treated in the ``magnetic channel'' via the ansatz $n_{i \uparrow} n_{i \downarrow}\rightarrow  \half N_i \sum_\sigma n_{i\sigma} + M_i S^z_i - \left(\frac{1}{4} N_i^2 - M_i^2\right) $, where $S^z_i = \half \left(n_{i \uparrow} - n_{i \downarrow} \right)$, $N_i = \mean{\sum_\sigma n_{i\sigma}}$ is the mean occupancy, and $M_i = \mean{S^z_i}$ is the local magnetization at site $i$. The quantities $N_i$ and $M_i$ are to be determined by enforcing the self consistency conditions of the mean field theory. Our detailed numerical calculation exploits translational symmetry along the nanoribbon (see \Fig{fig:maggs}(a)) where we have used up to 4000 equally spaced points to sample the 1D Brillouin zone. \Fig{fig:maggs} shows the results for the values of physical parameters typical for graphene. We find that there are two possible magnetic configurations where the moments along the two edges of nanoribbon are oppositely aligned (the anti-ferro (AF) configuration, see \Fig{fig:maggs}(a)) and aligned in the same direction (the ferro (FM) configuration, see \Fig{fig:maggs}(b)). For both these configurations, the moments are concentrated on the sites at the edge layers -- this point will be important in the discussion below. At half-filling (one electron per site), the ground state has an AF structure consistent with earlier results\cite{son2006}. With the doping of holes denoted by $\delta$ ($\delta$ is defined as the doping per edge atom of the ribbon), the ground state changes to the FM configuration at a critical doping\cite{macdonald2009}. This critical doping $\delta_c$  required for the first order transition is dependent on the width of the ribbon and we find that $\delta_c \sim \frac{a}{W}$ as shown in \Fig{fig:maggs}(c), the ``magnetic phase diagram''. In the remainder of the paper, we develop an analytical theory of the physics behind this phase diagram.

The continuum field theory\cite{Herbut2006} that captures the physics of the Hamiltonian \eqn{eqn:Hubbard} has the action ($\hbar = 1$)
\beq\mylabel{eqn:Action}
\begin{split}
{\cal S}[\Psi] & = \int \D{^{2+1}r} \,\,\, \sum_\sigma \Psi^\star_\sigma(r)\left(\OneFour \dou_\tau + {\cal H}_K \right) \Psi_\sigma(r) \\
& + Ue_0 a^2 \int \D{^{2+1}r} \left[\frac{1}{4}n(r)^2  - \{s^z(r)\}^2 \right]
\end{split}
\eeq
where $r = (\br, \tau)$, $\br \equiv (x,y)$ is the position vector where $x$-coordinate is along the length of the ribbon and $y$ along the width, $\tau$ is the imaginary time that runs from $0$ to $\beta$ (inverse temperature), $\Psi^\star_\sigma = \left(\psi^\star_{A+\sigma} \,\, \psi^\star_{B+\sigma} \,  \, \psi^\star_{A-\sigma} \,\, \psi^\star_{B-\sigma}  \right)$ is the array of Grassmann fields with $A/B$ and $+/-$ being, respectively, sublattice and valley indices, $n(r) = \sum_{a\nu\sigma} \psi^\star_{a\nu\sigma}(r) \psi_{a\nu\sigma}(r)$ is the number density, $s^z(r) = \half \sum_{a\nu\sigma} \sigma \psi^\star_{a\nu\sigma}(r) \psi_{a\nu\sigma}(r)$ is the spin density, $\OneFour$ is the $4\times4$ identity matrix, and
\beq\mylabel{eqn:Dirac}
{\cal H}_{K} =  v_F \left( 
\begin{array}{cc}
\btau \cdot \bP  & \bzero \\
\bzero & -\btau^* \cdot \bP \\
\end{array} 
\right) -\mu \OneFour
\eeq
with $v_F = e_0 a$, and $\bP = -i \grad$, the momentum operator, $\btau = \tau_x \be_x + \tau_y \be_y$ ($\tau_{x,y}$ -- Pauli matrices in the sublattice space, $\be_{x,y}$ are spatial basis vectors).  The action, written in a form that anticipates magnetism,  can be studied by introducing a Hubbard-Stratanovich field $m(r)$ to decouple the $(s^z)^2$ term in \eqn{eqn:Action}, while the $n^2$ term can be treated in a straightforward manner via a Hartree shift. The action becomes
\beq\mylabel{eqn:ActionPsiM}
\begin{split}
{\cal S}[\Psi,m] & = \int \D{^{2+1}r}  \sum_\sigma \Psi^\star_\sigma(r)\left(\OneFour \dou_\tau + {\cal H}_K \right)\Psi_\sigma(r) \\
& +  \int \D{^{2+1}r}  \left[ m(r) s^z(r) + \mbox{g.c.}  + \frac{1}{U e_0 a^2} |m(r)|^2 \right]
\end{split}
\eeq
which upon integration of the fermion fields yields
\beq\mylabel{eqn:ActionM}
{\cal S}[m] = \ln{\prod_\sigma \det\left(-G^{-1}_\sigma[m]\right)} + \frac{1}{U e_0 a^2} \int \D{^{2+1}r} |m(r)|^2 ,
\eeq
where the inverse Green's function is given by $
-G^{-1}_\sigma(r,r',m) = \left[\OneFour \dou_\tau + {\cal H}_K + \frac{\sigma}{2} \{m(r) + m^*(r)\} \OneFour\right] \delta(r - r')$. The magnetic ground state of the system can be described by a saddle point of the action \eqn{eqn:ActionM}. The saddle point field $m_s$ satisfies the condition
\beq\mylabel{eqn:SaddlePoint}
\frac{1}{U e_0 a^2} m_s(r) = \sum_\sigma \frac{\sigma}{2} G_\sigma(r,r;m_s).
\eeq

\begin{figure}
\centerline{\includegraphics[width=0.38\myfigwidth]{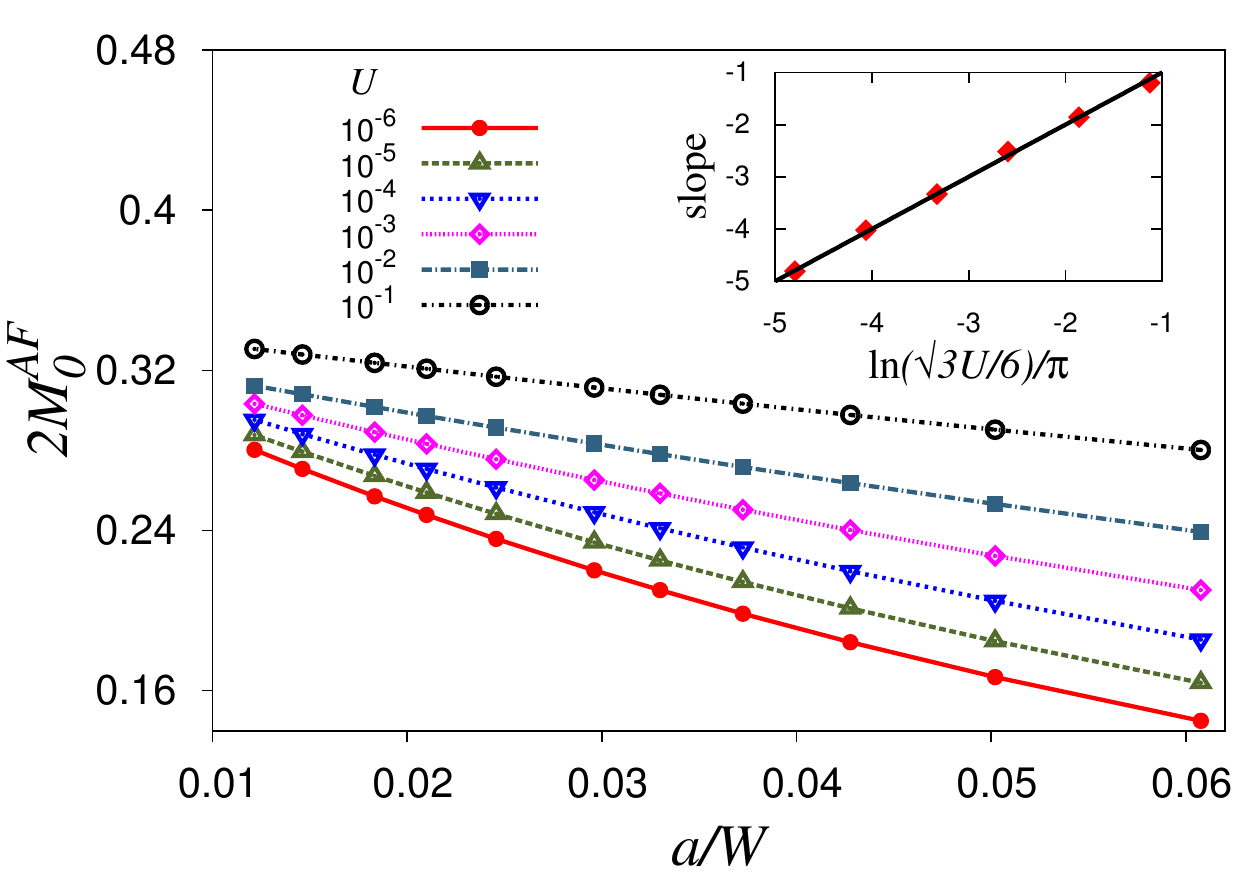}}
\caption{(color online) (a) Width dependence of magnetic moment of the AF configuration for  different 
values of $U$ obtained from the full numerical solution compared with the analytical result \eqn{eqn:MAF0}. Inset: The logarithmic dependence of the slope of $M^{AF}_0$ vs. $a/W$  on $U$ -- points are obtained from the numerical simulations, solid line is from \eqn{eqn:MAF0}. }
\mylabel{fig:AFcompare}
\end{figure}

We consider two different saddle point ansatzes for the undoped nanoribbons.  For the AF configuration, we have 
\beq\mylabel{eqn:AFansatz}
m_s^{AF}(y) = Ue_0 a^2 \left[\frac{M^{AF}}{a} \delta_D(y-W) - \frac{M^{AF}}{a} \delta_D(y) \right]
\eeq
where $M^{AF}$ is the moment associated with one edge, and FM configuration has 
\beq\mylabel{eqn:FMansatz}
m_s^{FM}(y) = U e_0 a^2 \left[\frac{M^{FM}}{a} \delta_D(y-W) + \frac{M^{FM}}{a} \delta_D(y) \right],
\eeq
where $M^{FM}$ is, again, the moment associated with one edge, and $\delta_D$ is the Dirac delta function. Solution of  $M^{AF}$ and $M^{FM}$ is aided by the observation that the right hand side of  \eqn{eqn:SaddlePoint} is the magnetic moment density response of a nanoribbon with applied edge Zeeman fields. As shown in ref.~\cite{bhowmick2010}, this response is highly nonlinear akin to that of sensory organs. Thus, for the AF configuration, \eqn{eqn:SaddlePoint} reduces to
\beq\mylabel{eqn:SelfCons}
M^{AF} = \half \left[ \frac{1}{3} + \frac{a}{\pi W} \ln\left(\sqrt{3} U M^{AF} \right) \right]
\eeq
For the FM configuration, $M^{FM}$ satisfies a similar equation sans  the factor of $\sqrt{3}$ in front of $U$. For wide ribbons $W \gg a$, we find
\begin{align}
M^{AF}_0 & \approx  \half \left[ \frac{1}{3} + \frac{a}{\pi W} \ln\left(\frac{\sqrt{3} U}{6} \right) \right] \mylabel{eqn:MAF0} \\
 M^{FM}_0 & \approx  \half \left[ \frac{1}{3} + \frac{a}{\pi W} \ln\left(\frac{U}{6} \right) \right] \mylabel{eqn:MFM0}
\end{align}
in undoped case (hence the subscript 0). We further find that the AF configuration has a one-particle excitation gap given by
\beq\mylabel{eqn:AFgap}
\frac{\varepsilon_g}{e_0} \approx - \frac{2 a}{W} \left(\frac{U}{6}\right) \ln\left(\frac{U}{6}\right)
\eeq
\Fig{fig:AFcompare} presents a comparison of the analytical results for the AF configuration with the full numerical calculations. We find excellent {\em quantitative} agreement of the calculated edge moment with the theory (\eqn{eqn:MAF0}) over five decades of $U$ (see inset\footnote{The slope in the  inset of \Fig{fig:AFcompare} is the coefficient of the linear term obtained by fitting a polynomial $M(a/W) = c_0 + c_1 (a/W) + c_2 (a/W)^2$.}  of \Fig{fig:AFcompare}). Similar quantitative agreement is found for the energy gap \eqn{eqn:AFgap}. We also find excellent quantitative agreement between our theory and the numerical calculations of the FM state. 

Our theory predicts the AF configuration to be the ground state of the undoped ribbons for any width. This owes to the fact that the edge moment $M^{AF}_0$ is larger than $M^{FM}_0$, 
\beq
\Delta M_0 = M^{AF}_0 - M^{FM}_0 = \frac{\ln\sqrt{3}}{2\pi} \frac{a}{W}
\eeq
which, remarkably, is independent of $U$. The difference in the ground state energies $(E_0^{AF} - E_0^{FM})$ per unit repeat distance along the length of the ribbon is estimated as $
\frac{\Delta E_0}{e_0} \approx -U\left[\frac{\ln \sqrt{3}}{3} \frac{a}{\pi W} + \left( \frac{(\ln \sqrt{3})^2}{2} + \frac{(6 - \pi) \ln \sqrt{3}}{6} \ln\left(\frac{U}{6} \right) \right) \left(\frac{a}{\pi W} \right)^2 \right]$, 
where the leading term in $a/W$ arises from the correlation energy (proportional to $M^2$), and the second term $\sim (a/W)^2$ also contains the kinetic energy of the electrons in the effective bands. The ground state of the undoped system is always the AF configuration the physics of which traces back to the sensory organ like response of zigzag-edge terminated ribbons.

\begin{figure}
\includegraphics[width=0.31\myfigwidth]{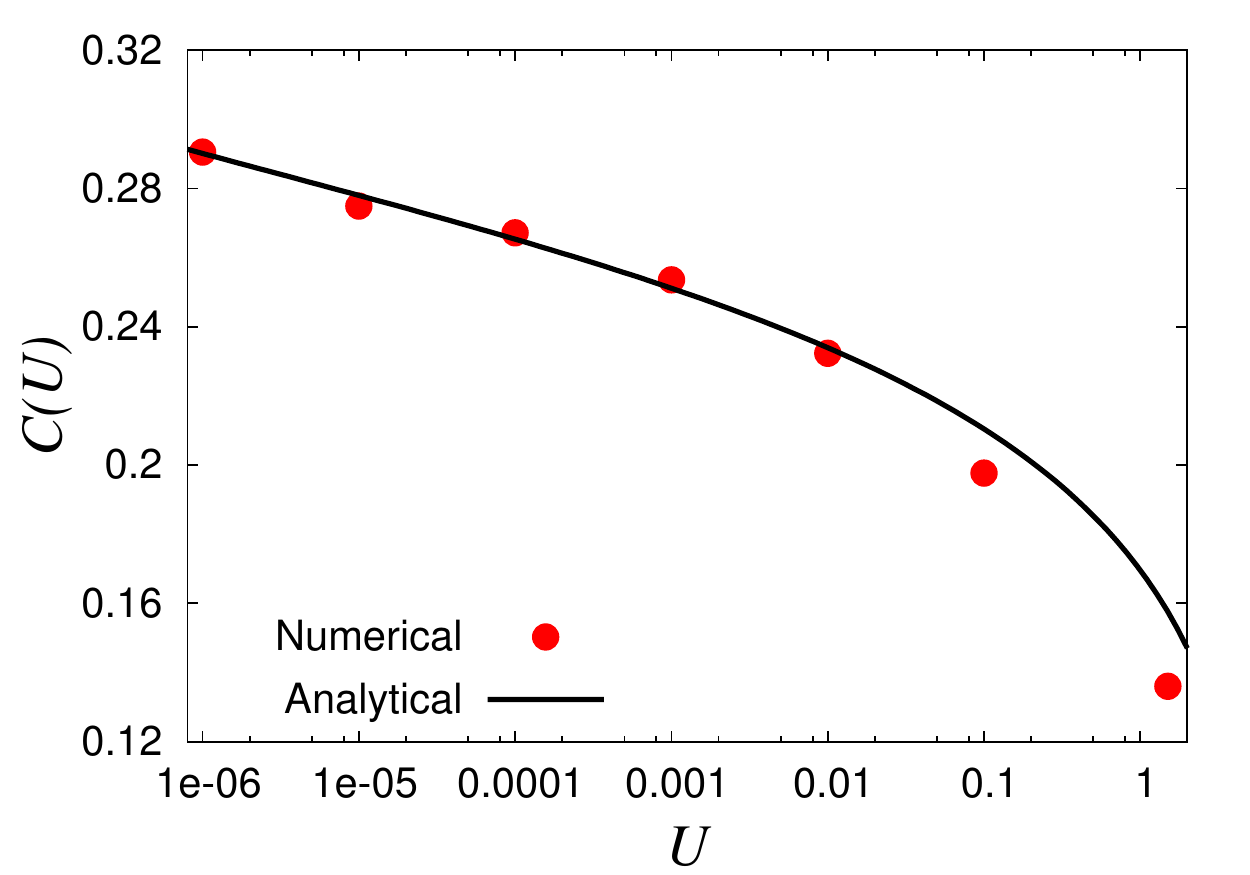}
\caption{(color online) Dependence of $C(U)$ on $U$. Critical doping $\delta_c$ for transition from AF to FM state is given by $\delta_c = C(U) \frac{a}{W}$ (see \eqn{eqn:DeltaC}).}
\label{fig:Udelc}
\end{figure}

Upon doping the system with holes ($\delta >0$) or electrons $(\delta < 0)$, the edge magnetization changes. For the AF configuration, we find
\beq\mylabel{eqn:MAFdelta}
M^{AF}(\delta) = M^{AF}_0 - \frac{|\delta|}{2}
\eeq
while, interestingly, for the FM configuration
\beq
\mylabel{eqn:MFMdelta}
M^{FM}(\delta) \approx M^{FM}_0 - s(U) \frac{W}{a} \delta^2
\eeq
where
$s(U) \approx \frac{\pi}{4} \left(1 + \frac{2}{\ln(U/6)} \right)$. This leads to an energy difference between the two states 
\beq
\Delta E(\delta) = -\frac{2 Ue_0}{3} \left( \Delta M_0 - \frac{|\delta|}{2} + s(U) \frac{W}{a} \delta^2 \right)
\eeq
to order $a/W$. We see that a first order transition from the AF to the FM configuration occurs at a critical doping 
\beq\mylabel{eqn:DeltaC}
\delta_c = C(U) \frac{a}{W},\;\; C(U) = \mbox{$\frac{1-\sqrt{1 - \ln3 \left(1 + \frac{2}{\ln(U/6)} \right)}}{\pi \left( 1 + \frac{2}{\ln(U/6)}\right)}$} 
\eeq
\Fig{fig:Udelc} shows a comparison of this analytical result \eqn{eqn:DeltaC} with the
numerical values of $C(U)$, and again, excellent quantitative
agreement is found over many decades of $U$. For larger values of $U$
(such as that found in graphene), we find quantitative agreement up to
about 10\%. This owes to the fact that  larger values of $U$ results in
a small contribution from the bulk states, which is not captured in our
edge mode based analytical theory. To the best of our knowledge, this is
the first analytical and quantitative theory of the magnetism in zigzag-edged
ribbons. An important point brought about by the theory is that
interpreting the magnetism in these systems via a Stoner-Landau theory
of the form \eqn{eqn:LandauTheory} may be too simplistic. Indeed, the
energy functional has a non-analytic structure $E(M) = A M^2 + B M \ln
 DM$ ($A,B,D$ constants), that has roots in the sensory organ like Weber-Fechner response of
zigzag-edge ribbons. 

It is important to ensure that quantum fluctuations does not change
the qualitative physics uncovered by the analytical theory. To this
end, we performed variational Monte Carlo calculations of the ground
state. Our trial wave function $\ket{\varPsi}= g^{D} \ket{FS,M}$, where
$D$ is the double occupancy operator, $g$ is a the Gutzwiller factor
that penalises double occupancy, $\ket{FS,M}$ is the filled Fermi sea
state constructed by imposing an edge magnetization $M$ on the edge
layers. For the AF configuration $M$ has opposite site on the two
edges while for the FM configuration $M$ is equal on both edges. The
optimal values of $g$ and $M$ are obtained so as to minimize the
ground state energy. We have studied ribbons with $W\approx 6 a$ with
$U=1.5$. We find that the ground state at zero doping is the AF
configuration, with an edge moment 0.20 which is expectedly
smaller than the mean field value 0.27 owing to quantum fluctuations. Furthermore, the ground state changes to FM
configuration at a critical doping $\delta$ of $0.025$. Both the AF and FM
edge moments are stable, in that they remain unchanged upon
increasing the length of the ribbon (for a given width $W$),
proving that the magnetism is robust. While these calculations are
prohibitively expensive for the determination the full phase diagram,
these results provide evidence for the correctness of our analytical
theory.

\begin{figure}
\centerline{\includegraphics[width=0.16\myfigwidth]{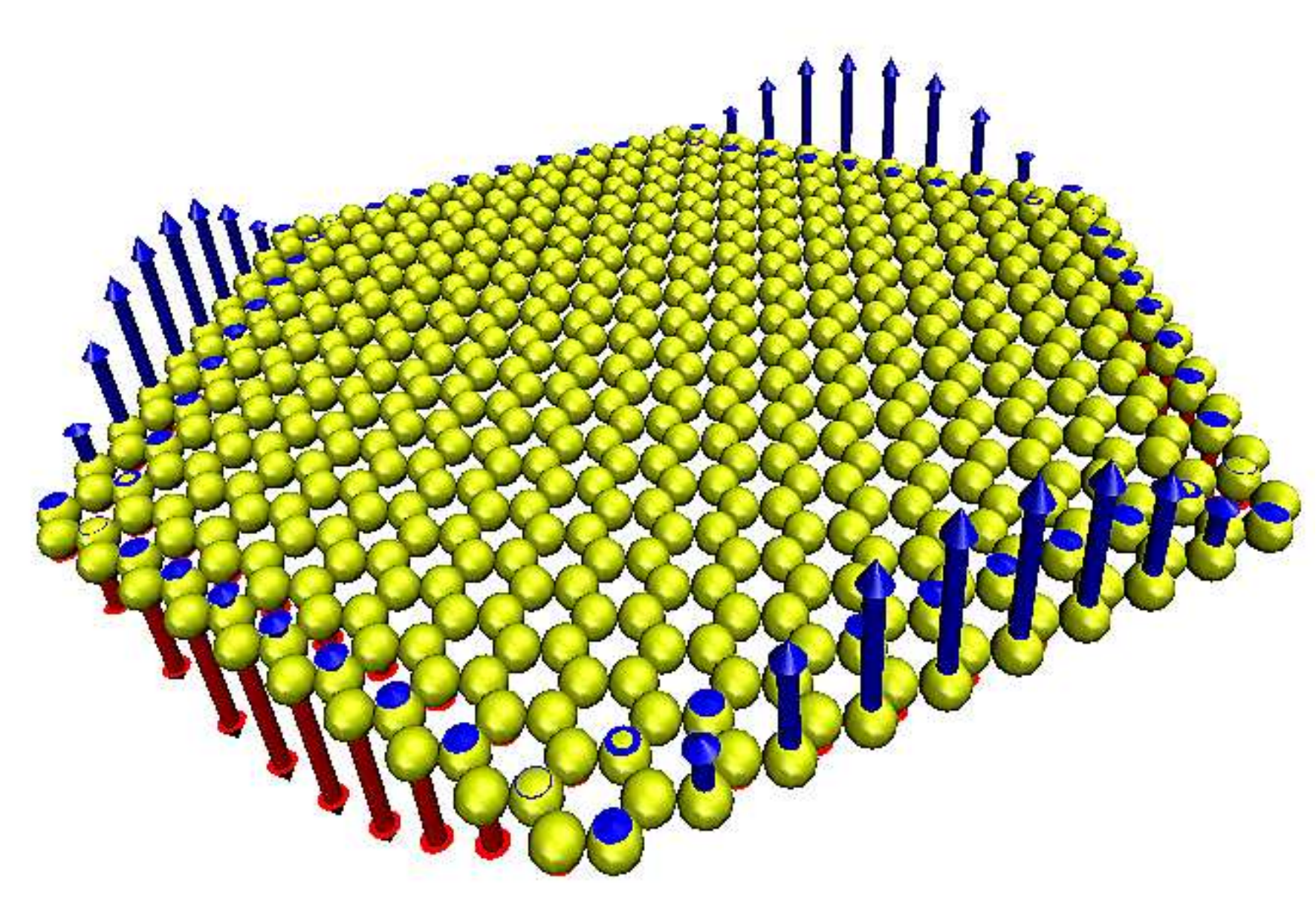}~~~~\includegraphics[width=0.15\myfigwidth]{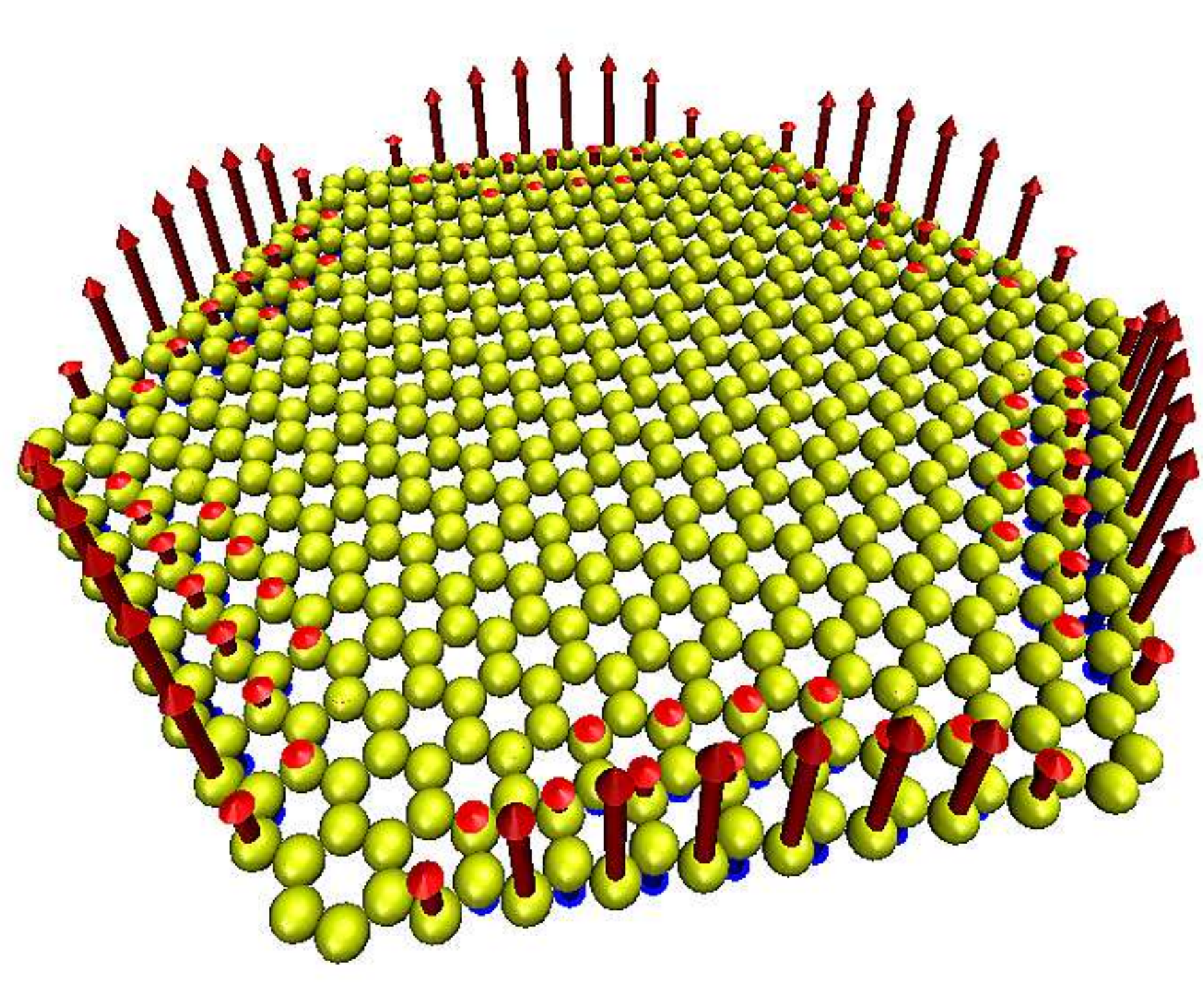} }
\centerline{(a)~~~~~~~~~~~~~~~~~~~~~~~~~~~~~~~~~~(b)}
\caption{(color online) (a) Mean-field AF configuration in a zigzag edge terminated hexgonal nanodot $(U=1.5)$ (b) FM configuration obtained upon doping ($\delta=0.1$).}
\label{fig:NS}
\end{figure}

\begin{figure}
\centerline{\includegraphics[width=0.20\myfigwidth]{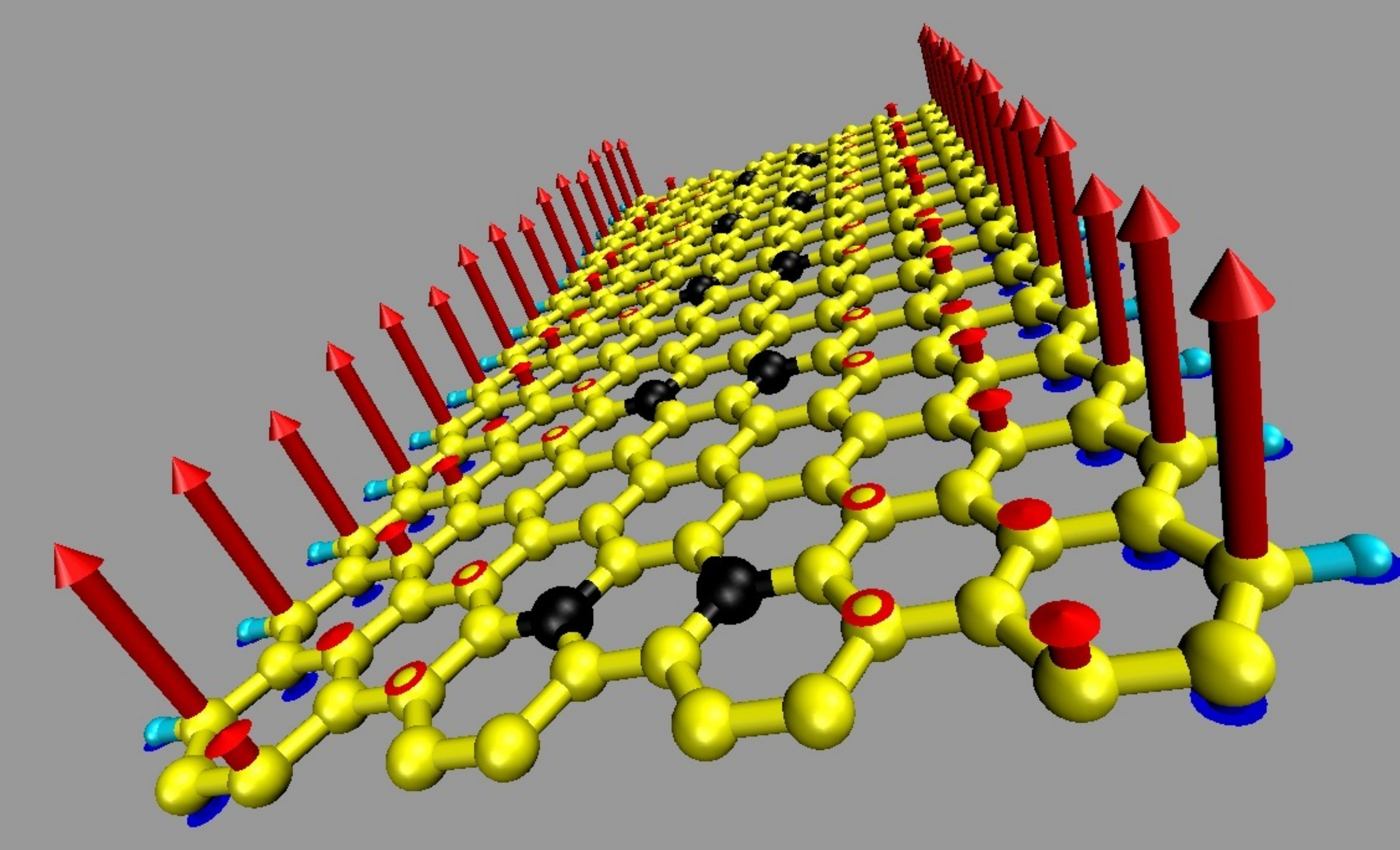} }
\caption{(color online) FM structure of the boron substituted zigzag edge terminated graphene nanoribbon obtained from first principles calculations. Dark atoms are borons and light ones are carbon. Hydrogen passivation of the edge atoms is also shown. }
\label{fig:Boron}
\end{figure}

We have further investigated other zig-zag edge terminated
nanostructures. For a undoped hexagonal nanodot, we find that the
magnetic configuration has an AF structure\cite{rossier2007} as shown
in \Fig{fig:NS}(a), which upon doping, changes to a FM
type (see \Fig{fig:NS}(b)).  Doping in practical applications can be
achieved by gating. We have also explored possible chemical
modification of graphene nanoribbons to produce an ``internal doping'' to engender the magnetic transition. \Fig{fig:Boron} shows a FM structure
of a zigzag edge terminated nanoribbon\footnote{These calculations were done using Quantum
Espresso code,\cite{QE-2009} using plane-wave basis set and ultrasoft pseudo-potential. The energy cutoff
for the plane-wave basis for wavefunctions is set to be 40 Ry. Electron exchange-correlation is treated with a local
density approximation(LDA, Perdew-Zunger functional). Nanoribbons are simulated using a supercell geometry, with a
vacuum layer of 15\AA~between any two periodic images of the ribbon. A k-point grid of 24x1x1 k points (periodic direction
of the ribbon along x-axis) is used for sampling Brillouin zone integrations for this geometry.} with boron atoms
substituted\cite{cnr2009} in place of carbons, while the undoped nanoribbon has
a AF configuration within the same calculation. As is evident, the
results of this paper suggest many interesting possibilities of using
zigzag edge terminated graphene nanostructures in applications.

AM thanks CPDF, IISc for support. VBS is grateful to DST for
Ramanunjan and MONAMI grants, and DAE-SRC for generous support. The
authors thank Umesh Waghmare for help with the first principles
calculations, R.~Shankar and Jayantha Vyasanakere for discussions.

\bibliography{ref}
\end{document}